\documentstyle[12pt,bezier,emlines]{article}
\title{\bf Coherent States of Two-dimensional 
 Josephson Junction Networks }
\author{ W. Krech\thanks{owk@rz.uni-jena.de}\hskip4mm and
K. Yu. Platov\thanks{okp@rz.uni-jena.de} \thanks{
Permanent address:
Laboratory of Cryoelectronics,
Physics Department, Moscow State University, 
Moscow 119899 Russia}\\
\sl Friedrich-Schiller-Universit"t Jena\\
\sl Physikalisch-Astronomisch-Technikwissenschaftliche Fakult"t\\
\sl Institut fr Festk"rperphysik\\
\sl 07743 Jena\\
\sl Max-Wien-Platz 1}
\begin{document}
\maketitle
\begin{abstract}

 We have investigated numerically  the  phase--locking
behavior  of two-dimensional Josephson junction arrays,
taking into account
a finite inductance ($l\stackrel{>}{\sim}1$) of the unit 
cell  and  external magnetic fields.
Within this  model we  
demonstrate the 
existence of multi--valued {\it I--V}
curves of the network. Different  branches  
of the {\it I--V} curve
correspond to  "in--phase" and "anti--phase"  
coherent states  with  high and low  
levels of output power, respectively. 
The arrays show remarkable hysteretic  transitions between 
these states under influence of external 
magnetic fluxes in the range 
$ - \Phi_0/4 < \Phi_{ext} < \Phi_0/4 $. 
\end{abstract}
\vfill
\eject

\section{   Introduction}
    Arrays of Josephson junctions have 
been  recognized as  remarkable candidates for
voltage-tunable high-frequency radiation 
sources for a long time \cite{Jain1,Hansen1}. 
During the last years different types 
of two-dimensional (2D) arrays  have been investigated both 
theoretically (see e.g. \cite{Lee1})
and experimentally \cite{Tinkham1}.
A number of experiments demonstrated 
coherent behavior of such systems. 
However,  the dependence of the power on 
the Josephson frequency  was different
from the one predicted for  ideal case of
a completely phase-locked system \cite{Benz1,Benz2}. 
The aim of this work is to analyze  possible reasons
for this discrepancy. We  demonstrate that such a
behavior can be explained by a finite inductance 
of the array unit cell. As a result, depending on the history,
the system possesses several coherent states with 
different frequencies for a given value of the bias current 
and therefore,  exhibits hysteretical 
features  in spite of the small value of 
the McCumber parameter ($\beta < 0.4$) . 
Moreover, the structure  shows no single-valued 
voltage-flux dependence.
This remarkable behavior is  similar to 
the dynamical properties  of  a many-junction 
SQUID \cite{Lewandowski1,Meyer1} recently investigated 
in detail \cite{Darula1}. All numerical results 
presented here have been obtained 
using the PSCAN network analyzer \cite{Odintsov1,Polonsky1}.

\section{ Hysteretical Behavior of Multi-Junction Interferometers}

In order to understand the behavior of  two--dimensional 
structures let us first have a look on a quasi-1D 
Josephson array biased by a DC current 
(so-called parallel multi--junction
interferometer) as shown in Fig. 1.
 To avoid additional self--field effects
due to the supplied current we feed separate 
bias currents in every branch  of the 
M--junction interferometer line. 
In the following section we discuss  the results 
of the dynamical simulations 
within the framework of  the RCSJ model.
The current through each junction is given by 
\vskip1mm
\begin{equation}
 \beta\ddot{\phi_k}+\dot{\phi_k}+\sin\phi_k={i_k},
\end{equation}
\centerline{$i_k=I_k/I_c$.}
\vskip5mm
Here the McCumber parameter $\beta $ is defined  as  
$\beta=2\pi I_cC{R_n}^2/\Phi_0$
where  $\phi_k$ is the Josephson phase difference  over the {\it k-th}
junction and
  $\Phi_0$ is the magnetic  flux quantum.
  $C$, $R_n$ and $I_c$ are the junction capacitance, 
normal resistance and critical current, respectively.
  $I_k$ is the total current flowing through the {\it k-th} junction.
Introducing  dimensionless parameters, we can formulate
current conservation relations and  flux-quantization 
conditions:
\vskip1mm
\begin{equation}
\phi_k-\phi_{k+1}={\phi}_{ext,k}+l_0i_{(k)},
\end{equation}
\begin{equation}
i_1=i_b-i_{(1)}, 
\end{equation}
\begin{equation}
  i_k=i_b+i_{(k-1)}-i_{(k)} \quad,  \quad 1<k<M,
\end{equation}
\begin{equation}
   i_M=i_{(M-1)}+i_b,
\end{equation}
where we defined the following normalized parameters:
$ l_0=2 \pi I_c L_0 / \Phi_0$,
${\phi}_{ext,k}=2\pi{\Phi}_{ext,k}/\Phi_0$. 
$ i_b=I_b/I_c $ is the 
supplied DC bias current,
$i_{(k)}=I_{(k)}/I_c$ is the current flowing horizontally 
through the respective inductance (cf. Fig. 1).
  $k$ is the junction  number ($k=1,...,M$),
  $L_0$ is the loop inductance.
  ${\Phi}_{ext,k}$  stands for the external magnetic flux.
It is useful to  define also a mutual phase difference 
for all junctions as 
\begin{equation}
 \theta_{lk}=\phi_l-\phi_k \quad , \quad 1 \le l,k \le M
\end{equation}
where the subscripts $k,l$ refer to the respective junction number.
The frequency of the junction oscillations is 
\begin{equation}
 \omega=\overline{v},
\end{equation}
here $\overline{v}$ indicates the normalized DC component 
of the voltage across the  junctions (We use the symbol 
$\overline{x}$ to denote the time  average value of a function $x$).
We restrict our simulations to the (quite realistic) case
of a uniform external magnetic field  
(${\phi}_{ext,k} = \phi_{ext}$ ) and  choose 
a small value of the McCumber parameter, $\beta \leq 0.4$,  
to obtain  non-hysteretical behavior of a single junction. 

It is known that for   $l_0\ll1$  a multi-junction 
interferometer shows a continuous and 
periodic dependence of the frequency   on external 
magnetic fields \cite{Likharev1}.
With increasing inductances the situation 
becomes more complex.  
When a  weakly coupled system with
$l_0>1$  enters the  narrow vicinity of 
the point $\phi_{ext}=0$, its behavior 
becomes hysteretical. Fig. 2 shows 
the Josephson frequency of the system 
as a function  of the external
magnetic flux. For vanishing magnetic field
the array  can exist in three different stable states
each of them  determined by the current 
circulating inside the loop. The first state is trivial and
no DC current circulates around the complete loop 
including the  junctions $J_1$ and $J_M$, i.e.
$ \overline{i_{(k)}}=0$, $\theta_{lk}=0$. 
The frequency  $\omega_0$ of this state is equal 
to that of a single
junction with bias current $i_b$.
We call this state $S_0$ ({\it zero}--state). 
The second state  $S_+$ ({\it plus}--state with 
clockwise circulating DC current) and the
third state  $S_-$ ({\it minus}--state with 
counterclockwise circulating DC 
current)
correspond to the presence of $\Phi_0$ 
 or $ - \Phi_0$ in the whole loop of the system, 
respectively. 
The latter two states have different phase relations 
between neighbors of the row ($\theta_{lk} \ne 0$). 
The situation is illustrated in Fig. 3.
We would like to point out  that even for a system with 
identical junctions ($i_{c,k} = i_c$)  
the frequencies $\omega_+$, $\omega_-$ 
of the states $S_+$, $S_-$ 
are different from the frequency  $\omega_0$ of the 
$S_0$ state  in the case of vanishing external flux,
$\phi_{ext}=0$.
A spread of the critical currents causes a small shift  
of the frequency--flux relationship.
As a result the frequencies $\omega_-$ and 
$\omega_+$ become different  even in  zero magnetic 
field. 
Simulations show that for 
$\phi_{ext}/\phi_0 \ll 1 $ the state $S_0$  
is unstable and  will be replaced by either 
the state $S_-$  or $S_+$  after some time  
(However, this time may be long  compared 
to the period $2\pi \omega^{-1}$ of the 
Josephson oscillation).

\section{ Coherent States of Simple 2D Arrays (N = 2) }
 A  simple example of a 2D array is the system 
shown in Fig. 4
(Following tradition, M denotes the width 
of the array and N the number of rows).
The circuit consists of 
two multi-junction interferometers 
($M = 4,\quad N = 2$) connected in series.  
Commonly,  elementary cells of a 2D array
have additional junctions in the horizontal branches.  
In the case of undercritical currents flowing in 
the horizontal branches these junctions can 
be replaced by linearized
(sin$\phi \approx \phi$) 
inductances $L_J=\Phi_0/2\pi I_c$.
Therefore, the loop inductance can not 
be lower than $2L_J$.
In the following we only consider  
arrays without junctions 
in the horizontal branches  (i.e., the 
parameter $l_0$ can be lower than 1). 
Furthermore, no DC 
voltages drop over the horizontal superconducting 
inductances and therefore  the frequencies of the 
Josephson oscillations of all junctions along the row
are equal. This is the so-called low frequency ({\it lf})
mechanism of synchronization. 
The coherent state of the whole system is
due to the high--frequency current  ({\it hf})
flowing  through 
the external load with impedance 
$z(\omega)=r_s+j \omega l_s$. 

We define the spread of critical currents as
$\delta i_c =$ max $ \mid i_{c,k}-i_{c,l}\mid $ 
and the corresponding  mean value by
$\overline{i_c} = \sum_{k}^{}i_{c,k}/MN$.
For simplicity, we assume the identity of all 
products  $V_{c,k}=I_{c,k}R_{n,k}$.
Furtheremore, we suppose that
the  load matching relation $R_s \approx R_{array}$ 
are being satisfied. In order to detect whether complete locking  
takes place we calculate output power released 
by the system to the external load. In  
the  uniform mode  
the output  power delivered to a matched load is given by 
\begin{equation}
P_{out}=MNP_1
\end{equation}
where $P_1={I_c}^2R_n/8$.

The results of numerical
simulation are shown in Figs. 5a-c. Let us now compare the 
ideal case (all junctions are identical)
with the nonideal case (spread of the critical currents).
In the ideal case one finds that the 
output power per junction as predicted 
by the common RCSJ model for a single junction 
over all the current range of interest.
A system with nonidentical critical currents 
exhibits hysteretical behavior and  
within a certain range of the bias current
the system can be in two states. 
In order to explain  such a behavior let us 
consider the following qualitative description.  
From the results  of Sec. 1
one can conclude that the practically uniform 
locking is represented by a situation
where both interferometers of the system  are 
in the $S_0$ state. In this case
no DC currents are flowing in the horizontal branches, 
and all junctions  oscillate almost in phase. 
Otherwise, nonidentical critical currents result in 
DC currents flowing in the horizontal branches. 
We call them "excess currents". They are  
caused by extracting bias currents 
from junctions with smaller values of $i_c$ and
injecting them into junctions with larger $i_c$. 
In this way the {\it lf} interaction redistributes
the currents to match the voltages across the rows.
Fig. 6 helps clarify the situation.
Note that we have to take into account  
the inductances in all  branches of 
the unit cell. This is important  because all 
loops of the array with the 
same geometrical length must have  the same 
inductance and consequently the same quantization 
condition. 
The values of the excess currents in the horizontal 
lines are different from each other
due to the non-identity of the junctions. 
 Flowing through inductances,  
excess currents produce additional mutual phase shifts 
in a cell  and  influence 
the system in  the same way as the external 
magnetic field. As a consequence the 
 phase differences for all 
junctions  do not vanish, $\theta_{lk} \ne 0 $,
 the {\it hf} current and the according oscillation output 
power  decrease. 
   The excess current for every pair ($k,l$) of non-identical 
junctions can be estimated to
\begin{equation}
\Delta i_{exc,kl} \approx i_b(i_{c,k}-i_{c,l})/\,\overline{i_c}
\end{equation} 
where  $i_{c,k}$ and $i_{c,l}$  are the critical currents of
the {\it k-}th and the {\it l-}th junctions in the row.
These currents produce an additional phase shift
\begin{equation}
\Delta \phi \approx l_0i_b/4\sum_{l,k}^M(i_{c,k}-i_{c,l})
(l-k)/\,\overline{i_c}
\end{equation} 
which is proportional to the bias current.
The value of the {\it hf} current  flowing through the 
external load  for the phase-locking state is 
\begin{equation} \label{Ss}
i_{\omega}=\frac{Nv_{\omega}}{z_{ext}( \omega )+Nz_{cell}( \omega )/M }
\end{equation}
where
\begin{equation}
 z_{cell}( \omega )=j \omega l_0/4+r_n/(j \omega cr_n+1) \quad,
\end{equation}
\begin{equation}
      z_{ext}( \omega )=j \omega l_s+r_s \quad ,
\end{equation}
\begin{equation}
 v_{\omega} \approx 2 \overline{v} / 
(({{\overline{v}}^2+1)}^{1/2}+ \overline{v} ) \quad . 
\end{equation}
Fig. 7 presents  the corresponding 
dependence of the  {\it hf} current 
on the frequency  for two different values of $l_s$.
The  amplitude of the {\it hf} current
remains constant or  decreases slowly for
$\omega \ge {\omega}_c $,  
 depending on the load inductance 
(For the time being we suppose that 
the value of $r_s$ remains constant).
 Assuming  $\theta_{lk}=0 $ 
and neglecting therefore 
a decrease of the {\it hf} current due to the non-zero phase 
[150qshift between the voltage oscillations of the respective 
junctions, we obtain this well-known result.
Taking into account 
$\theta_{lk} \ne 0$  as a result of 
{\it lf} interaction the current $i_{\omega}$ 
in Eq. \ref{Ss}
decreases  more rapidly with frequency. 
On the other hand, the strength of {\it lf} interaction
increases as $i_b$. 
The result of the competition between these currents
determines both the phase distribution $\theta_{lk}$ 
and the oscillation frequency of the array (see Fig. 8).
The phase-locking state with  
high output power will break down if the 
{\it lf} locking  strength overcomes
that of {\it hf} interaction. 
After switching one of the rows into   
the $S_-$ state and the other one into the $S_+$ state, 
the system remains in a locking state 
but the new phase relations result in a small output power. 
The flux quantization condition for the boundary 
loop (the junctions $J_1,J_5,J_4,J_8$)
is still $\Delta \phi=0$.
The switching back can not be achieved 
by a small decrease of the bias current because the amplitude 
of the {\it hf} current decreased radically after 
the switching thus making the {\it hf} interaction 
strength  very small. 
The switching back corresponds to an attempt 
of the system  to change its  $S_+S_-$ state into a  
 $S_-S_+$ state. During this transition both rows 
unavoidably pass the $S_0$ state and 
the {\it hf} interaction
restores the uniform phase-locking again. 
This qualitative explanation is supported by Fig. 9  
which shows the output power delivered by 
the system as  a function of $l_0$
for several values of the bias current $i_b$. 
Note that 
for $l_0\stackrel{<}{\sim}1$ only the $S_0$ state exists. 
On the other hand,
with increasing bias current  the strength of 
the {\it lf} interaction also increases  and 
practically excludes the 
 $S_0$ state for all $l_0 > 1$. Vertical lines 
in Fig. 9 show 
the boundaries of the regions where 
 both states are possible.
In the case $l_0>1$ 
the system has 
different oscillation frequencies. 
The state with high output power can be reached  only  
when the electromagnetic coupling due to the external 
load overcomes the strenght of the {\it lf} interaction.

\section{ Dynamical Properties of 2D Arrays ( $N > 2$ )}
  We have verified the features discussed above 
by performing simulations on more complex arrays of
different sizes ($N \le 4$, $M \le 6$). 
We summarize the results as follows:

(i) If all the rows have the same mean value 
of the critical current and 
$\delta I_c \le 0.03I_c$, then the behavior 
of the array does 
not depend on $l_0$ and the system behaves as
a single junction with a value of the
critical current $MI_c$
and normal resistance $NR_n/M$.  
  
(ii) If  all the  rows have the same mean value 
of the critical current and 
$ 0.03I_c < \delta I_c \le 0.15I_c$, then the 
behavior of the whole array  strongly depends 
on the value of $l_0$. In the range $l_0 \le 1.0$ the 
array can have one coherent state only. 
Otherwise, in the region $l_0 > 1.0$, the features 
of the system are determined by  a three-state 
dynamics. 
Using an optimization procedure with respect to
the  load inductance, phase-locking  
can be achieved for  given  values of
$l_0$ and  $R_{array}\approx R_s $ 
over a wide frequency range 
$0< \omega < 2 \omega_c$. 
A hysteresis exists in the current-voltage 
characteristic of  the system as well 
as in the voltage-flux  dependence for
$\omega > 2 \omega_c$. 

(iii) If the mean values of the critical 
currents  of  all the rows are different and  
the condition $l_0 \le 1.0$ 
is satisfied then the main features (i.e. stability and
strength) of the synchronization are 
not qualitatively different from those of an 
1D structure (see e.g. \cite{Jain1}).

(iv) In the opposite case ($l_0 > 1.0$)
the main difference is 
that the presence of the external 
field can form a coherent state of the 
system, even if it does not exist for an initial
state  with $\phi_{ext}=0$ (see Figs. 10a,b).
The restriction of the value of $ \delta I_c $ 
is more tight than  in the case (ii).
The coherent state of the system breaks up 
for any value of the  external load and of 
the unit cell inductance if  $\delta I_c \ge 0.1I_c$. 
\section { Conclusions }
We have performed numerical simulations of the behaviour 
of arrays of overdamped  Josephson junctions in 
magnetic fields.
We have shown that the phase-locking  of 2D Josephson 
junction arrays strongly depends  on the value of 
the  unit cell inductance $l_0$. An array with inductance 
$l_0>1 $ and with  a critical current spread 
$\delta I_c \stackrel{>}{\sim} 0.03I_c$ may show  several 
coherent states with different frequencies 
for a specific value of the bias current. This leads to 
a hysteresis in the {\it I-V} curve of the network. 
A magnetic field can lead  either to  a
destruction of a coherent state or to 
transitions  between coherent states 
with strongly different radiation power levels.
\section*{Acknowledgments}
The authors  would like to express their 
thanks to  BMFT and DAAD  for supporting this work. 
\eject


                            Figure Captiones

\centerline{FIGURE CAPTIONS}
\vskip1mm
{\small Fig. 1. Multi-junction interferometer.}
\vskip1mm
{\small Fig. 2. The frequencies $\omega_0$,
$\omega_-$ and  $\omega_+$ of a M-junction 
interferometer as  function of the 
external  magnetic flux $\phi_{ext}$.
The solid line shows the frequency dependence 
for the {\it zero}-state $S_0$. Parameters: $ M=6$, $i_b=1.3$, $l_0=6.0$. }
\vskip1mm
{\small Fig. 3. The phase relations 
$\theta_{1k}=\theta_1-\theta_k $ for 
a M-junction interferometer 
as function of the external magnetic flux
$\phi_{ext}$. Parameters: $M=6$, $i_b=1.3$, $l_0=6.0$}.
\vskip1mm
{\small Fig. 4. A pair 
of  current biased four-junction  interferometers 
connected in series  with an external load $R_s$, $L_s$ as 
an example of a simple 2D array. }
\vskip1mm
{\small Fig. 5a. $I-V$ characteristic of a 2D structure 
with  $l_0$=3.0 for the spread $\delta i_c=0$ (solid line) and 
$\delta i_c=0.1i_c$.}
All the junctions are overdamped ($\beta=0.4$).
The latter curve has a lower branch corresponding to 
an {\it in}-phase state ($\omega = \omega_0$) and  an 
upper branch corresponding to an {\it anti}-phase state.}
\vskip1mm

{\small Fig. 5b. Output power per junction 
of the 2D structure with $l_0$=3.0  
as  function of the bias current $i_b$ for
$\delta i_c=0$ (solid line) and 
$\delta i_c=0.1i_c$. }
\vskip1mm
{\small Fig. 5c. Phase relations 
$\theta_{31},\theta_{35}$  for the combinations $J_3 - J_1$ 
and $J_3 - J_5$ as  function of the bias current $i_b $ for 
$\delta i_c=0.1 \overline{i_c}$  (for $l_0$=3.0)}.
\vskip1mm
{\small Fig. 6. Current configuration in the 2D structure. 
Horizontal and vertical arrows indicate the direction 
of the "excess current".}
\vskip1mm
{\small Fig. 7. Current-frequency relationship
of Eq. (11) for two values of the external load 
inductance ($l_s=5.0$ and $l_s=3.0$).}
\vskip1mm
{\small Fig. 8.  $I-V$ curves for two values of 
the external load inductance $l_s=5.0$ and $l_s=3.0$.}
\vskip1mm
{\small Fig. 9. Output power per  junction 
of the {\it in-} and {\it anti-}phase 
states as  function of the unit cell inductance 
for several values of the bias current
($i_b=1.3, 1.5, 1.8$) and a fixed 
value of the external impedance. Vertical 
lines are guides to the eye.}
\vskip1mm
{\small Fig. 10. Frequencies  of the rows 
of the 2D array (M=6, N=4, $l_0=1.5$, $\delta i_c=0.08 \overline{i_c}$ )  
as  function  of $\phi_{ext}$ for different values
of the bias current $i_{b}=1.4$ (a) and $i_{b}=2.0$ (b).
An external flux $\phi_{ext} \ge 0.09\phi_0$
is responsible for the coherent state.}

\end{document}


                            Figure 1.

section 1 of uuencode 4.13 of file 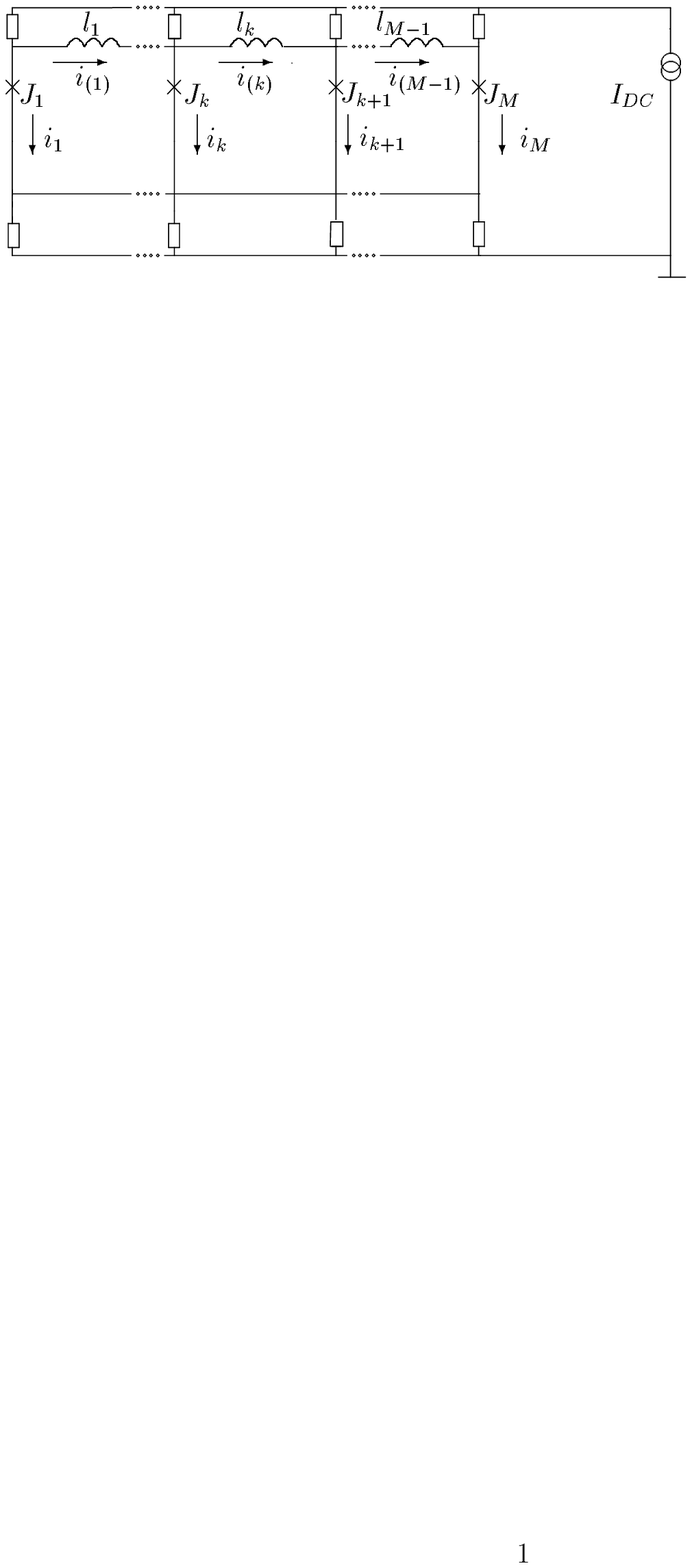    by R.E.M.